# Meta-Gratings: Beyond the Limits of Graded Metasurfaces for Wavefront Control

Younes Ra'di, Dimitrios L. Sounas, and Andrea Alù[*]

*Department of Electrical and Computer Engineering, The University of Texas at Austin, Austin, TX 78712, USA*

*To whom correspondence should be addressed: alu@mail.utexas.edu

*Graded metasurfaces exploit the local momentum imparted by an impedance gradient to transform the impinging wavefront. This approach suffers from fundamental limits on the overall conversion efficiency and it is challenged by fabrication limitations on the maximum spatial resolution. Here, we introduce the concept of meta-gratings, which enables arbitrary wavefront engineering with unitary efficiency and significantly lower demands on the fabrication precision. We employ this concept to design reflective metasurfaces for wavefront steering without limitations on efficiency. A similar approach is envisioned for transmitted beams and for arbitrary wavefront transformation, opening promising opportunities for highly-efficient metasurfaces for extreme wave manipulation.*

Metasurfaces are thin structured arrays that have attracted significant attention for the level of control of electromagnetic waves that they enable [1]-[6]. Phase-gradient metasurfaces, in particular, have recently been explored to tailor the electromagnetic wavefront in unprecedented ways to realize low-profile lenses, holograms, beam steerers and other ultrathin optical devices.

The vast majority of these geometries have been based on the generalized laws of reflection and refraction [7], based on which we can engineer the local reflection and/or transmission coefficients so that the impinging wave acquires the tangential momentum necessary to be locally rerouted towards the desired direction [7]-[17]. As research in this area has progressed, a few papers [18]-[20] have pointed out how this approach is inherently limited in efficiency, especially when extreme wave manipulation is considered. For instance, in the canonical problem of plane wave steering, it has been realized that the phase-gradient approach fundamentally suffers from a tradeoff between efficiency and steering angle. Since this design method does not take into account the impedance mismatch between impinging and desired wavefronts, a dramatic reduction of efficiency and an increase of scattering to spurious directions is expected as the steering angle increases, even in the ideal limit in which we are able to engineer the desired phase profile with infinite resolution. Considering also fabrication limitations, these inefficiencies grow even larger. These issues have important implications in the design of metasurfaces for extreme wave transformations, beyond what commonly achievable with conventional gratings [21]-[22].

These inherent limitations of phase-gradient metasurfaces can be overcome by designing metasurfaces with a surface impedance profile determined by the exact boundary conditions for the local impinging and scattered electromagnetic fields, which include impedance matching considerations. Based on this approach, however, transforming an incident wave towards an arbitrary direction with unitary efficiency in general requires locally active or strongly non-local metasurfaces [18]-[19]. Using the interference with auxiliary evanescent modes [23] or leaky modes [24], one can engineer passive impedance profiles that emulate the required non-local effects and get close to unitary efficiency. Such metasurfaces, however, are still subject to implementation difficulties when the limited fabrication resolution are considered, since they also require the discretization of a continuous, fastly varying, impedance profile.

Here, we introduce a different route to the design of metasurfaces for arbitrary wavefront transformation, which we call *meta-gratings*, enabling unitary efficiency without the need of deeply subwavelength elements, turning away from the discretization of a continuous phase or impedance profile. We exploit the well-established physics of a periodic grating, suitably tailoring the local period to select a discrete set of diffraction orders, and enrich its wave control by introducing a complex metamaterial scatterer within each unit cell. As we prove in the following, the scatterer can be designed to suppress undesired diffraction orders, and reroute the incident power towards the desired ones with unitary efficiency. Since we do not focus on discretizing a continuous phase/impedance profile, the metasurface does not require subwavelength resolution, and the typical sizes involved are in the order of the wavelength of excitation.

In the following, we focus on the challenging problem of beam steering in reflection with unitary efficiency, which is not possible without relying on active or nonlocal metasurfaces if we consider continuous impedance profiles [18]. Interestingly, we rigorously prove that a suitably tailored individual scatterer per unit cell is sufficient to enable highly complex diffraction scenarios of significant practical interest, with unitary efficiency. We then explore the physical mechanism behind this anomalous diffraction, and discuss the extension of the proposed approach to other scenarios of practical interest.

Consider first the basic geometry of an array of horizontally oriented magnetic dipoles located at a distance $h$ from a ground plane (see Fig. 1a). We assume that the structure is illuminated with a transverse magnetic (TM) plane wave propagating in the $yz-$ plane with electric field $\mathbf{E}_{\text{inc}} = E_0 (\mathbf{y} \cos\theta_{\text{inc}} + \mathbf{z} \sin\theta_{\text{inc}}) \exp(-jk_0 \sin\theta_{\text{inc}} y + jk_0 \cos\theta_{\text{inc}} z)$, where $E_0$ is its amplitude, $k_0$ is the free-space wavenumber and $\theta_{\text{inc}}$ is the incidence angle defined in the counterclockwise direction with respect to the $z$–axis. Due to the magnetic dipoles, the metasurface sustains an induced surface magnetic current $\mathbf{J}_m(x,y,z) = \mathbf{z} I_m^x \delta(z-h) \sum_{m=-\infty}^{\infty} \sum_{n=-\infty}^{\infty} \delta(x-ma)\delta(y-nb)\exp(-jk_{y0}nb)$, where $k_{y0} = k_0 \sin\theta_{\text{inc}}$, $\delta(x)$ is the Dirac's delta function, $I_m^x$ is the magnetic current on each dipole, and $a$, $b$ are the periodicities in the $x$ and $y$ directions, respectively. $I_m^x$ is related to the magnetic field $H_{\text{ext}}^x$ in the absence of the array as $I_m^x = j\omega \hat{\alpha}_{mm}^{xx} H_{\text{ext}}^x$, where $\hat{\alpha}_{mm}^{xx}$ is the effective magnetic polarizability of the particles in the presence of all other particles. In absence of Ohmic loss or gain, the quantity $\hat{\alpha}_{mm}^{xx}$ satisfies the passivity condition [25]

$$\text{Im}\left\{\frac{1}{\hat{\alpha}_{mm}^{xx}}\right\} = \frac{\omega}{\eta_0 ab} \sum_{n \in propagating} \frac{1}{\cos\theta_{0n}} \cos^2(k_{z0n} h), \quad (1)$$

which stems from the fact that the radiated power extracted by the array equals the radiated one. Here, $k_{z0n} = k_0 \cos\theta_{0n}$ is the wavenumber in the $z$–direction of the $0n$ mode (the first and second indices are the Floquet orders with respect to the $x$– and $y$–axes, respectively), $\eta_0$ is the free-space wave impedance, and $\theta_{0n}$ is the corresponding direction angle defined in the clockwise direction with respect to the $z$–axis. Depending on the period, the power can be coupled to a discrete number of diffraction orders [26], and here we select the period so that no higher-order modes can propagate due to the periodicity in the $x$–direction. The radiated fields for the $0n$ mode can be written as $\mathbf{E}_{0n} = (\mathbf{y} - \mathbf{z} k_{y0n}/k_{z0n}) Q_{0n} \exp(-jk_{y0n}y - jk_{z0n}z)$, where $Q_{0n} = I_m^x \cos(k_{z0n}h)/(ab)$ and $k_{y0n} = k_{y0} + 2n\pi/b = k_0 \sin\theta_{0n}$ is the wavenumber in the $y-$ direction.

In order to fully transmit the incident wave to a direction different than the specular one, we design the period to align one of the higher-order Floquet modes to this desired angle. In order for all the incident power to be transferred to the $0(-1)$ mode, for instance, we need to make sure that the structure does not radiate power into the specular channel $00$ and all other propagating higher-order channels. The direct reflection from the ground plane can be cancelled by the $00$ Floquet mode, which is achieved if $Q_{00} = E_0 \cos\theta_{inc}$, leading to the condition

$$\frac{1}{\hat{\alpha}_{mm}^{xx}} = j\omega \frac{2}{\eta_0 ab} \frac{1}{\cos\theta_{00}} \cos^2(k_{z00}h). \tag{2}$$

The required magnetic polarizability is purely imaginary, showing that cancellation of specular reflection happens at the meta-grating resonance. Comparing this equation with Eq. (1), we obtain the design equation

$$\frac{1}{\cos\theta_{00}} \cos^2(k_{z00}h) = \frac{1}{\cos\theta_{0(-1)}} \cos^2(k_{z0(-1)}h), \tag{3}$$

from which we can calculate the required distance of the magnetic dipoles from the ground plane. In addition to the cancellation of the fundamental mode, we need to make sure that all higher-order modes other than $0(-1)$ do not carry power away from the metasurface. This may be achieved by making sure that these modes are evanescent, which happens within the following range of incident and diffraction angles

$$\begin{cases} \theta_{0(-1)} < \arcsin\left(2\sin\theta_{\text{inc}} - 1\right), & 0 < \theta_{\text{inc}} < \arcsin\left(\dfrac{1}{3}\right) \\ \theta_{0(-1)} < \arcsin\left(\dfrac{1}{2}\sin\theta_{\text{inc}} - \dfrac{1}{2}\right), & \arcsin\left(\dfrac{1}{3}\right) < \theta_{\text{inc}} < \dfrac{\pi}{2} \end{cases}. \tag{4}$$

Figs. 1b shows the difference between right- and left- hand sides of Eq. (3) for different values of $h$ and $\theta_{0(-1)}$ where $\theta_{\text{inc}} = 30$ deg. A zero corresponds to a solution of Eq. (3), for which the optimum polarizability can be found using Eq. (2). Interestingly, if the array of dipoles is located right on top of the ground plane ($h=0$) the only available solution is retro-reflection, i.e., $\theta_{0(-1)} = -\theta_{\text{inc}}$. Retroreflection with unitary efficiency can also be achieved for any other value of height $h$. As an example, in Fig. 1b we show a metasurface consisting of an array of magnetic dipoles located on the ground plane that provides unitary retroreflection (design point indicated by a star). These results are in agreement with recent works demonstrating retroreflection with large efficiency using passive metasurfaces [27]-[30], yet here we realize this effect with a single resonant inclusion per unit cell.

Figs. 1b-c show that this basic geometry is also capable of rerouting all the incident power to arbitrary desired directions by controlling the height $h$, with large flexibility, independent of the incidence angle. As an extreme example, we show in Fig. 1c an array of dipoles that reroutes an incident wave with $\theta_{\text{inc}} = 10$ deg to $\theta_{0(-1)} = -77$ deg (red star). The upper and lower right subsets present the near-field distributions around this array, and the power coupled into modes 00 and $0(-1)$, respectively, showing that the illuminated power can be entirely reflected back into the $0(-1)$ mode with unitary efficiency at the design frequency. By simply tailoring the distance from the ground plane, it is possible to reroute the incident wave into an extremely slanted direction, without having to worry about complex, fastly varying phase profiles. In the results shown in Fig.

1, we used a wire loop with a gap (detailed geometry described in [25]), however a similar response can be achieved with other geometries offering a magnetic response, for instance with dielectric or metallic-dielectric nanoparticles at optical frequencies [31]-[33]. It is remarkable to notice how a single particle per unit cell, suitably placed at a specific distance from the reflector, can couple all the impinging energy to the desired, extreme angle. At a small distance above the particle, the local field impedance is exactly the one necessary to transform the impinging wavefront to the desired one, but this can be achieved with a totally passive single inclusion, exploiting the near-field interactions with the impinging wave. The upper right insets in Figs. 1b and 1c show the frequency response of this array, ensuring a moderately broadband response, considering the extreme wave manipulation for which the metasurface was designed. Less extreme beam steering enables broader frequency responses. The reason for this moderate broadband behavior is that the particle does not need to be deeply subwavelength, instead its volume can be fit within the total area of the unit cell without significant space constraints. This can relax the requirements on Q-factor for the particle, which is typically a significant challenge in conventional metasurface approaches. In addition, the single particle is slightly detuned from its individual resonance, due to the array coupling, as Eqs. (1)-(2) refer to the effective polarizability of the array, including the coupling with all other particles, which is not negligible especially as we design the particle with a size comparable to the meta-grating period.

Up to this point, we studied the basic geometry of an array of magnetic dipoles located over a ground plane. In this case, we are limited in terms of $\theta_{\text{inc}}$ and $\theta_{0(-1)}$ by Eq. (4), which ensures that all Floquet modes with $n \neq \{0, -1\}$ are evanescent. The white areas in Figs. 1b and 1c correspond to these forbidden regions. In addition, when both 01 and $0(-1)$ modes are propagating, this array does not have enough degrees of freedom to cancel one of these modes and reroute the entire incident wave to the other one. For example, with such a structure we cannot achieve perfect diffraction efficiency for a normally incident wave ($\theta_{\text{inc}} = 0$) to another arbitrary angle, since in this scenario both the $0(-1)$ and $01$ Floquet modes are propagating. Physically this is consistent with the fact that magnetic dipoles alone cannot break the symmetry of the array for normal incidence, implying that the $0(-1)$ and $01$ modes would be excited with equal amplitude. In order to reroute the entire illuminated wave into one of these degenerate modes and cancel out the fields in the other one, we need inclusions that scatter asymmetrically for normal incidence, which may be achieved with bianisotropic particles with coupled electric and magnetic responses [34], as in Fig. 2a. Such particles provide the symmetry breaking required to suppress the additional propagating Floquet mode.

The unwanted diffraction orders can be suppressed as long as the array polarizabilities satisfy [25]

$$\frac{1}{\hat{\alpha}_{\text{ee}}^{zz}} = j\omega \frac{2\eta_0}{3ab} \sin^2 \theta_{0(-1)} \cos^2 (k_0 h),$$

$$\frac{1}{\hat{\alpha}_{\text{mm}}^{xx}} = j\omega \frac{2}{\eta_0 ab} \cos^2 (k_0 h), \quad (5)$$

$$\frac{1}{\hat{\alpha}_{\text{em}}^{zx}} = -j\omega \frac{2}{ab} \sin \theta_{01} \cos^2 (k_0 h).$$

These conditions, combined with the passivity condition [25] $\cos^2(k_0 h) = \dfrac{4}{\cos \theta_{01}} \cos^2(k_{z01} h)$ and with the requirement that $0n$ higher-order Floquet modes with $|n| \geq 2$ are evanescent, enable steering a normally incident beam to an arbitrary angle $\theta_{01}$ larger than 30 deg. For $\theta_{01} < 30\,\text{deg}$, Floquet modes with $|n| \geq 2$ also carry power, generally reducing the efficiency. In this case, more degrees of freedom, e.g., a second inclusion in each unit cell, would be required to achieve unitary efficiency.

As an example, we have designed a metasurface based on omega bianisotropic particles that fully reflects a normally incident wave to $\theta_{01} = 83$ deg. An initial estimation for the design parameters can be obtained from Eq. (5) [see solutions in Fig. 2b]. These parameters were further optimized to achieve the desired functionality for a realistic bianisotropic particle in the form of an omega particle [34], with geometry detailed in [25]. Upper and lower insets in Fig. 2b present the frequency response of the structure and the fields at the design frequency. All the illuminated power is transferred to the 01 harmonic, a functionality that would be impossible with conventional metasurface approaches. At optical frequencies, similar scattering responses may be achieved with nanoparticles with broken inversion symmetry [35]. In this extreme case of wave rerouting, while the efficiency is unitary, the bandwidth of operation is quite narrow. For less extreme designs, or using more than one particle per unit cell, the bandwidth can be broadened.

More in general, here we have discussed the concept of meta-grating, based on which we can use a conventional diffraction grating approach to select the desired channels in reflection, and then tailor the electromagnetic response of an individual complex inclusion within each unit cell to realize unitary efficiency for arbitrary wavefront manipulation in a simple design that requires low spatial fabrication resolution. While here we focused on the case of reflective metasurfaces, the same concept can be extended to ideally manipulate transmitted wavefronts. For instance, by realizing two metasurfaces back to back with suitably tailored polarizabilities, it is possible to arbitrarily control reflection and transmission with unitary efficiency. Beyond beam-steering, we envision that the proposed concept will enable the realization of ultralow-profile, highly efficient planar lenses with short focal distances, all-angle absorbing or reflecting metasurfaces, holograms, and other wave transformations. Following this approach, we can create a large surface in which the period is locally modified to steer the beam in different directions, realizing focusing or holograms. A related approach has been recently followed to realize an efficient ultrathin focusing lens [36]. Our paper proves that it is not necessary to aim at implementing a local continuous phase and/or impedance profile within each unit cell of the metasurface to realize highly efficient wavefront manipulation. Instead, a single scatterer per unit cell is sufficient to realize arbitrary wave manipulation with efficiencies larger than what allowed with conventional metasurface approaches. The use of multiple inclusions within each unit cell may be exploited to increase the bandwidth of operation, or to realize multi-frequency, dual-polarization, or low-aberration responses [37]-[38].

This work was supported by the Department of Defense and the Welch Foundation.

**References**

[1] C. L. Holloway, E. F. Kuester, J. A. Gordon, J. O'Hara, J. Booth, and D. R. Smith, *An Overview of the Theory and Applications of Metasurfaces: The Two-Dimensional Equivalents of Metamaterials*, IEEE Antennas Propag. Mag. 54, 10 (2012).

[2] C. Pfeiffer and A. Grbic, *Metamaterial Huygens' Surfaces: Tailoring Wave Fronts with Reflectionless Sheets*, Phys. Rev. Lett. 110, 197401 (2013).

[3] Y. Zhao, X. -X. Liu, and A. Alù, *Recent Advances on Optical Metasurfaces*, J. Opt. 16, 123001 (2014).

[4] N. Yu and F. Capasso, *Flat Optics with Designer Metasurfaces*, Nat. Mater. 13, 139 (2014).

[5] S. A. Tretyakov, *Metasurfaces for general transformations of electromagnetic fields*, Philos. Trans. R. Soc. London A 373, 20140362 (2015).

[6] S. B. Glybovski, S. A. Tretyakov, P. A. Belov, Y. S. Kivshar, and C. R. Simovski, *Metasurfaces: From Microwaves to Visible*, Phys. Rep. 634, 1 (2016).

[7] Na. Yu, P. Genevet, M. A. Kats, F. Aieta, J. -P. Tetienne, F. Capasso, and Z. Gaburro, *Light Propagation with Phase Discontinuities: Generalized Laws of Reflection and Refraction*, Science 334, 333 (2011).

[8] A. V. Kildishev, A. Boltasseva, and V. M. Shalaev, *Planar Photonics with Metasurfaces*, Science 339, 1232009 (2013).

[9] S. Sun, K.-Y. Yang, C.-M. Wang, T.-K. Juan, W. T. Chen, C. Y. Liao, Q. He, S. Xiao, W.-T. Kung, G.-Y. Guo, L. Zhou, and D. P. Tsai, *High-Efficiency Broadband Anomalous Reflection by Gradient Meta-Surfaces*, Nano Lett. 12, 6223 (2012).

[10] F. Monticone, N. M. Estakhri, and A. Alù, *Full Control of Nanoscale Optical Transmission with a Composite Metascreen*, Phys. Rev. Lett. 110, 203903 (2013).

[11] A. Pors, M. G. Nielsen, R. L. Eriksen, and S. I. Bozhevolnyi, *Broadband Focusing Flat Mirrors Based on Plasmonic Gradient Metasurfaces*, Nano Lett. 13, 829 (2013).

[12] M. Esfandyarpour, E. C. Garnett, Y. Cui, M. D. McGehee, and M. L. Brongersma, *Metamaterial Mirrors in Optoelectronic Devices*, Nat. Nanotechnol. 9, 542 (2014).

[13] M. Kim, A. M. H. Wong, and G. V. Eleftheriades, *Optical Huygens Metasurfaces with Independent Control of the Magnitude and Phase of the Local Reflection Coefficients*, Phys. Rev. X 4, 041042 (2014).

[14] Z. Bomzon, G. Biener, V. Kleiner, and E. Hasman, *Space-Variant Pancharatnam–Berry Phase Optical Elements with Computer-Generated Subwavelength Gratings*, Opt. Lett. 27, 1141 (2002).

[15] E. Hasman, V. Kleiner, G. Biener, and A. Niv, *Polarization Dependent Focusing Lens by Use of Quantized Pancharatnam–Berry Phase Diffractive Optics*, Appl. Phys. Lett. 82, 328 (2003).

[16] E. Hasman, Z. Bomzon, A. Niv, G. Biener, and V Kleiner, *Polarization Beam-Splitters and Optical Switches Based on Space-Variant Computer-Generated Subwavelength Quasi-Periodic Structures*, Opt. Commun. 209, 45 (2002).

[17] V. S. Asadchy, Y. Ra'di, J. Vehmas, and S. A. Tretyakov, *Functional Metamirros Using Bianisotropic Elements*, Phys. Rev. Lett. 114, 095503 (2015).

[18] N. Mohammadi Estakhri, and A. Alù, *Wavefront Transformation with Gradient Metasurfaces*, Phys. Rev. X 6, 041008 (2016).

[19] V. S. Asadchy, M. Albooyeh, S. N. Tcvetkova, A. D´ıaz-Rubio, Y. Ra'di, and S. A. Tretyakov, *Perfect Control of Reflection and Refraction Using Spatially Dispersive Metasurfaces*, Phys. Rev. B 94, 075142 (2016).

[20] A. Epstein and G. V. Eleftheriades, *Huygens' Metasurfaces via the Equivalence Principle: Design and Applications*, JOSA B 33, A31 (2016).

[21] E. Silberstein, P. Lalanne, J.-P. Hugonin, and Q. Cao, *Use of Grating Theories in Integrated Optics*, JOSA A 18, 2865 (2001).

[22] P. Lalanne and P. Chavel, *Metalenses at Visible Wavelengths: Past, Present, Perspectives*, arXiv: 1610.02507 [physics.optics].

[23] A. Epstein and G. V. Eleftheriades, *Synthesis of Passive Lossless Metasurfaces Using Auxiliary Fields for Reflectionless Beam Splitting and Perfect Reflection*, Phys. Rev. Lett. 117, 256103 (2016).

[24] A. Díaz-Rubio, V. S. Asadchy, A. Elsakka, and S. A. Tretyakov, *From the Generalized Reflection Law to the Realization of Perfect Anomalous Reflectors*, arXiv: 1610.04780 [physics.optics].

[25] See the Supplemental Material

[26] A. K. Bhattacharyya, *Phased Array Antennas: Floquet Analysis, Synthesis, BFNs and Active Array Systems* (Wiley, Hoboken, NJ, 2006).

[27] Z.-L. Deng, S. Zhang, and G. P. Wang, *A Facile Grating Approach Towards Broadband, Wide-Angle and High-Efficiency Holographic Metasurfaces*, Nanoscale 8, 1588 (2016).

[28] X. Su, Z. Wei, C. Wu, Y. Long, and H. Li, *Negative Reflection from Metal/Graphene Plasmonic Gratings*, Opt. Lett. 41, 348 (2016).

[29] N. Mohammadi Estakhri, V. Neder, M. W. Knight, A. Polman, and A. Alù, *Visible Light, Wide-Angle Graded Metasurface for Back Reflection*, ACS Photonics 4, 228 (2017).

[30] D. L. Sounas, N. M. Estakhri, and A. Alù, *Metasurfaces with Engineered Reflection and Transmission: Optimal Designs Through Coupled-Mode Analysis*, in 10th International Congress on Advanced Electro-magnetic Materials in Microwaves and Optics - Metamaterials 2016, Crete, Greece, Sept. 2016.

[31] J. C. Ginn, I. Brener, D. W. Peters, J. R. Wendt, J. O. Stevens, P. F. Hines, L. I. Basilio, L. K. Warne, J. F. Ihlefeld, P. G. Clem, and M. B. Sinclair, *Realizing Optical Magnetism from Dielectric Metamaterials*, Phys. Rev. Lett. 108, 097402 (2012).

[32] M. R. Shcherbakov, D. N. Neshev, B. Hopkins, A. S. Shorokhov, I. Staude, E. V. Melik-Gaykazyan, M. Decker, A. A. Ezhov, A. E. Miroshnichenko, I. Brener, A. A. Fedyanin, and Y. S. Kivshar, *Enhanced Third-Harmonic Generation in Silicon Nanoparticles Driven by Magnetic Response*, Nano Lett. 14, 6488 (2014).

[33] Y. Ra'di, V. S. Asadchy, S. U. Kosulnikov, M. M. Omelyanovich, D. Morits, A. V. Osipov, C. R. Simovski, and S. A. Tretyakov, *Full Light Absorption in Single Arrays of Spherical Nanoparticles*, ACS Photonics 2, 653 (2015).

[34] A. N. Serdyukov, I. V. Semchenko, S. A. Tretyakov, A. Sihvola, *Electromagnetics of Bi-Anisotropic Materials: Theory and Applications* (Gordon and Breach Science Publishers, Amsterdam, 2001).

[35] R. Alaee, M. Albooyeh, M. Yazdi, N. Komjani, C. Simovski, F. Lederer, and C. Rockstuhl, *Magnetoelectric Coupling in Nonidentical Plasmonic Nanoparticles: Theory and Applications*, Phys. Rev. B 91, 115119 (2015).

[36] R. Paniagua-Dominguez, Y. F. Yu, E. Khaidarov, R. M. Bakker, X. Liang, Y. H. Fu, and A. I. Kuznetsov, *A Metalens with Near-Unity Numerical Aperture*, arXiv: 1705.00895 [physics.optics].

[37] F. Aieta, P. Genevet, M. A. Kats, N. Yu, R. Blanchard, Z. Gaburro, and F. Capasso, *Aberration-Free Ultrathin Flat Lenses and Axicons at Telecom Wavelengths Based on Plasmonic Metasurfaces*, Nano Lett. 12, 4932 (2012).

[38] F. Aieta, M. A. Kats, P. Genevet, and F. Capasso, *Multiwavelength Achromatic Metasurfaces by Dispersive Phase Compensation*, Science 347, 1342 (2015).

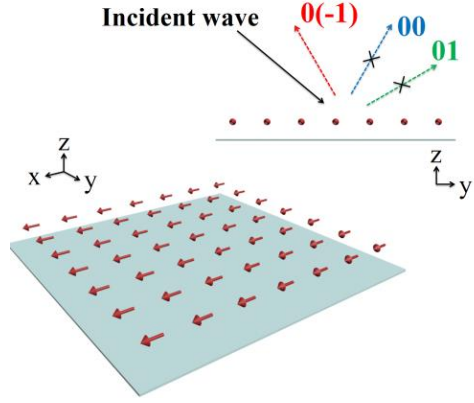

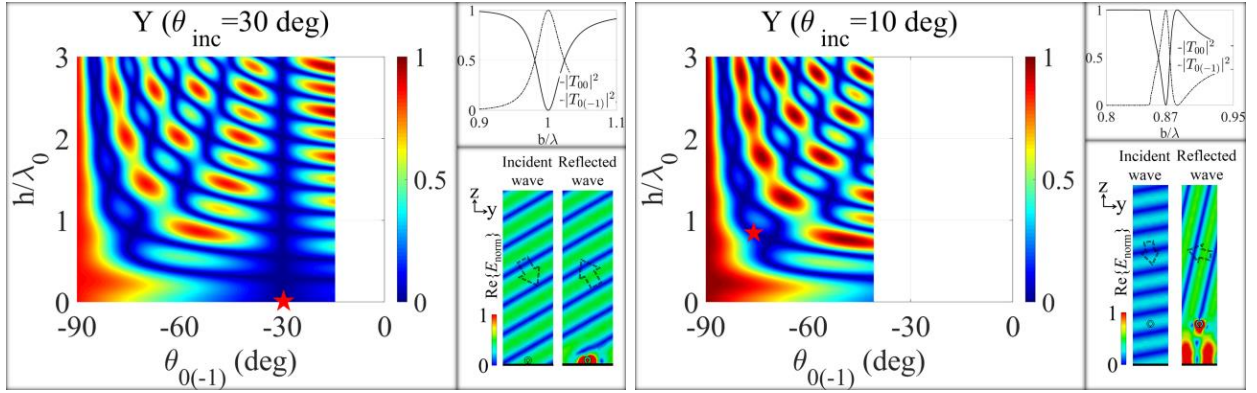

Fig.1. (a) Array of horizontal magnetic dipoles located over a ground plane. (b) Magnitude of $Y = \cos\theta_{0(-1)} \cos^2(k_{z00}h) - \cos\theta_{00} \cos^2(k_{z0(-1)}h)$, whose zeros correspond to the solutions of Eq. (3), for different design parameters, and $\theta_{inc} = 30$ deg. Dark blue areas correspond to parameters that satisfy Eq. (3), white areas to regions that support propagating Floquet modes other than $0(-1)$. (c) Solutions of Eq. (3) for $\theta_{inc} = 10$ deg. In panels (b-c), the red stars indicate the design parameters for the insets. Upper right insets show the reflected power into $00$ and $0(-1)$ Floquet modes versus normalized frequency. Lower right insets show the distribution of incident and reflected electric fields at the design frequency.

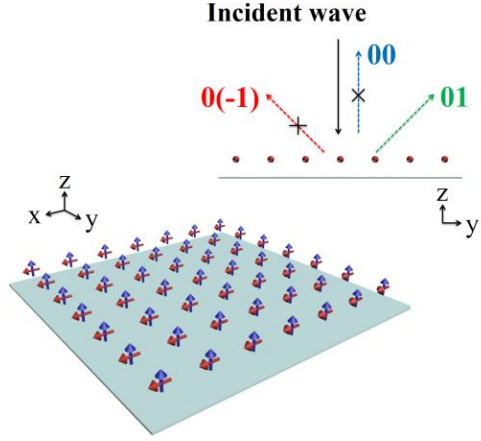

(a)

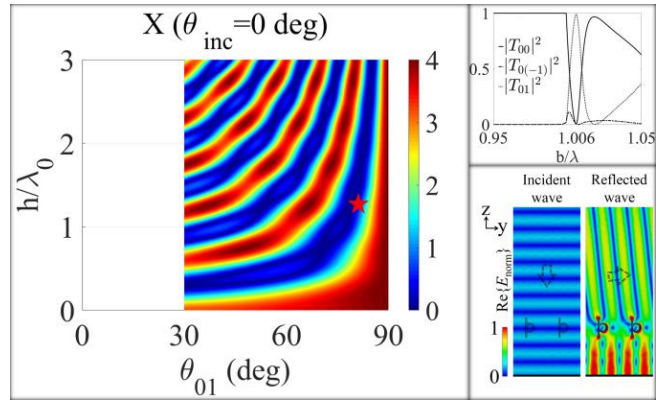

(b)

Fig.2. (a) Array of bianisotropic omega particles located over a ground plane. (b) Similar to Fig. 1, magnitude of $X = \cos\theta_{01} \cos^2(k_0 h) - 4\cos^2(k_{z01} h)$ for $\theta_{\text{inc}} = 0$ deg. The upper right inset shows the reflected power into $00$, $0(-1)$, and $01$ Floquet modes versus normalized frequency for the parameters indicated by a red star in panel b. The lower right inset shows the distribution of the incident and reflected electric fields at the design frequency.